\documentclass[12pt,preprint]{aastex}
\usepackage{graphics}

\def\gtorder{\mathrel{\raise.3ex\hbox{$>$}\mkern-14mu
             \lower0.6ex\hbox{$\sim$}}}
\def\ltorder{\mathrel{\raise.3ex\hbox{$<$}\mkern-14mu
             \lower0.6ex\hbox{$\sim$}}}

\shorttitle{Notch Filter Mask Performance}
\shortauthors{Debes et al.}

\begin{document}
\title{Using Notch Filter Masks For High Contrast Imaging of Extrasolar Planets}
\author{John H. Debes\altaffilmark{1},Jian Ge\altaffilmark{1},
Marc J. Kuchner\altaffilmark{2}}
\author{Michael Rogosky\altaffilmark{3}}

\altaffiltext{1}{Department of Astronomy \& Astrophysics, Pennsylvania State
University, University Park, PA 16802}
\altaffiltext{2}{Hubble Fellow, Department of Astrophysical Sciences, Princeton University, Princeton, NJ 08544}
\altaffiltext{3}{Pennsylvania State University Nanofabrication Facility,
      University Park, PA 16802}

\begin{abstract}
We present the first laboratory experiments using a notch-filter mask, a coronagraphic
image mask that can produce infinite dynamic range in an ideal Lyot coronagraph according to scalar diffraction theory.  
We fabricated the first notch-filter mask prototype with .25 $\micron$ precision
 using an e-beam lithography machine.  Our initial optical tests show that the prototype masks
generate contrast levels of 10$^{-5}$ at 3$\lambda/D$ and 10$^{-6}$ at $\sim 8 \lambda/D$, with
 a throughput of 27\%.  We speculate on the ``as-is'' performance of such a mask in the Hubble Space Telescope.
\end{abstract}   

\keywords{circumstellar matter---instrumentation: adaptive optics---methods:laboratory---planetary systems}

\section{Introduction}
\label{s1}
Directly imaging extrasolar terrestrial planets in reflected light requires
facing the extremely high predicted contrast ratios between planets and their host stars, e.g., $\sim 10^{-10}$
for an Earth analog orbiting a solar type star at quadrature. 
A planet-finding coronagraph must realize this contrast within a few diffraction widths
($\lambda/D$, where $\lambda$ is the wavelength of light, and $D$ is the long axis of the primary mirror) of the stellar image.  Though several 
coronagraph designs can achieve this contrast according to scalar
diffraction theory \citep{ks03}, substantial work
on mask design and laboratory investigation probably remains before this contrast can be achieved in practice.

Some coronagraph designs use image-plane masks to absorb on-axis light
and diffract it away \citep{malbet96,makidon01,kuchner02}.  Other designs use shaped or apodized pupils which benefit from
combining aperture shape and the pupil intensity distribution to remove
the wings of a circular aperture's PSF \citep{kasdin1,kasdin2,debes1,debes2,jian1}.  Image masks offer the
advantages that they explicitly remove starlight from the beam, and that
they can provide high contrast at small angles from the optical axis, given
sufficient control over low-spatial frequency modes.

Scattered light, wavefront errors, and mask 
construction errors can all degrade the contrast of a coronagraph.  For example, for any coronagraphic image mask, mid-spatial frequency
intensity errors near the center of the mask must be $\lesssim 10^{-9}$ \citep{kuchner02}.  Some of these errors can be controlled using active optics,
but these corrections will necessarily apply only over a limited range
of wavelengths.

Notch-filter masks offer a promising choice for planet-finding coronagraphs
\citep{kuchner02b}.
These image masks absorb most of the light from an on-axis
point source, and diffract all of the remainder onto a matched Lyot stop.
While Lyot coronagraphs with Gaussian image masks must
have a throughput of $\lesssim 1/2$ to reach 10$^{-10}$ contrast, linear
notch-filter masks have unlimited dynamic range according to scalar diffraction theory, and a throughput of $\sim(1-\epsilon)$, where $\epsilon$ is typically $\sim 0.3-0.5$.

Other coronagraph designs besides notch filter masks can create
perfect subtraction of on-axis light.  However, those designs
based on masks with odd symmetry \citep{rouan2,rouan1} or
interferometrically synthesized
masks with odd symmetry \citep{baudoz1, baudoz2} create nulls
that degrade as $\theta^2$, where $\theta$ is the angle from
the optical axis.  This rapid degradation means
that the finite size of a real star causes the coronagraph to
leak light at levels unsuitable for terrestrial planet detection.
Other designs, like the dual phase
coronagraphic mask with an apodized entrance pupil
\citep{soummer03a,soummer03b}, produce the needed null depth,
but must use masks with special chromatic behavior and require
precise, achromatic aperture apodization.
Notch-filter masks are intrinsically achromatic and like
the dual phase coronagraph, they create
nulls of order $\theta^4$ or slower \citep{kuchner02}.

Notch-filter masks can be designed such that they are binary--regions of the
mask are either opaque or transparent.  This is a great advantage as 
intensity errors are not an issue so long as the mask is sufficiently opaque,
leading to manufacturing constraints that are orders of magnitude smaller. 
However, the shape of the mask must be precisely 
reproduced, to the level of $\lambda f_\#$/3600 for a contrast of $10^{-10}$ within the search area.  For an instrument with $f/100$ and working 
$\lambda\sim.66\micron$,
this corresponds to a tolerance on the order of 20 nm.  Nanofabrication
techniques are required to reach this precision.

As part of a joint university-industry study partly funded by Ball Aeorospace and in collaboration with the Penn State Nanofabrication facility (Nanofab),
we have fabricated a prototype notch-filter mask and tested it in an 
experimental setup.  We discuss briefly the mask fabrication process in Section
 \ref{s2}, describe methods for modeling performance in
Section \ref{s2a}, review the experiments and results
 in Section \ref{s3}, and
discuss ways of improving performance in Section \ref{s4}.
\section{ Mask Design and Fabrication }
\label{s2}
Following the prescription laid out in \citet{kuchner02b}, we designed a notch-filter mask 
based on a $1-\mbox{sinc}^2$ band limited function.  The mask is comprised
of a vertically repeating pattern of opaque curves described by:
\begin{equation}
\label{eq:mbl}
\hat{M}_{BL}(x)=\pm .5\frac{\lambda_{min}}{D}\left(1-\mbox{sinc}^2\left(\frac{\pi \epsilon x D}{2 \lambda_{max}}\right)\right)
\end{equation}
where $\epsilon$ is used to determine the half power of the mask, the effective
distance where a companion could reasonably be detected; $\lambda_{min}$ and $\lambda_{max}$ are the minimum and maximum wavelengths of the spectral
band of interest.  In reality the mask 
is constructed and therefore sampled at some resolution, i.e. with a 
nanofabrication tool, such that the band limited equation is modified slightly;
\begin{equation}
\label{eqn:msamp}
\hat{M}_{samp}=\Pi \left(\frac{x}{w}-n\right) \star \hat{M}_{BL}(n)
\end{equation} 
where $n$ is over all integers, $w$ is the resolution of the
tool, and $\Pi$ is the tophat function.  The final
step is to convolve $\hat{M}_{samp}$ with a series of $\delta$ functions
spaced by $\lambda/D$ to produce the mask function:
\begin{equation}
\label{eqn:mnotch}
\hat{M}_{notch}=\sum^{N}_{k=0}\delta\left(y-\frac{k\lambda}{D}\right)\star \hat{M}_{samp}
\end{equation}
Repeating
the band limited curve on scales $\ltorder \lambda/D$ ensures that the
notch-filter mask becomes a good approximation of a transmissive mask
for spatial frequencies in the pupil plane $< D/\lambda$

We designed the mask for a f/158 system and a working wavelength of .633~$\micron$.
  Our choice of $\epsilon$ was .46, to allow an inner working distance
of 2-3~$\lambda/D$.  
For our working wavelength, the maximum size is $\sim$100~\micron.  While the
theory prescribes that the width of the stripes be no wider than 
$\lambda_{min} f_{\#}$, there is no prohibition from making the width smaller
and so we built the mask with stripes of width 25~$\micron$.
There are two advantages to undersizing, one is guarding against blue light
leakage in a broadband case and the other is allowing future tests to be
performed at smaller $f_{\#}$.  
 
\begin{figure}
\plotone{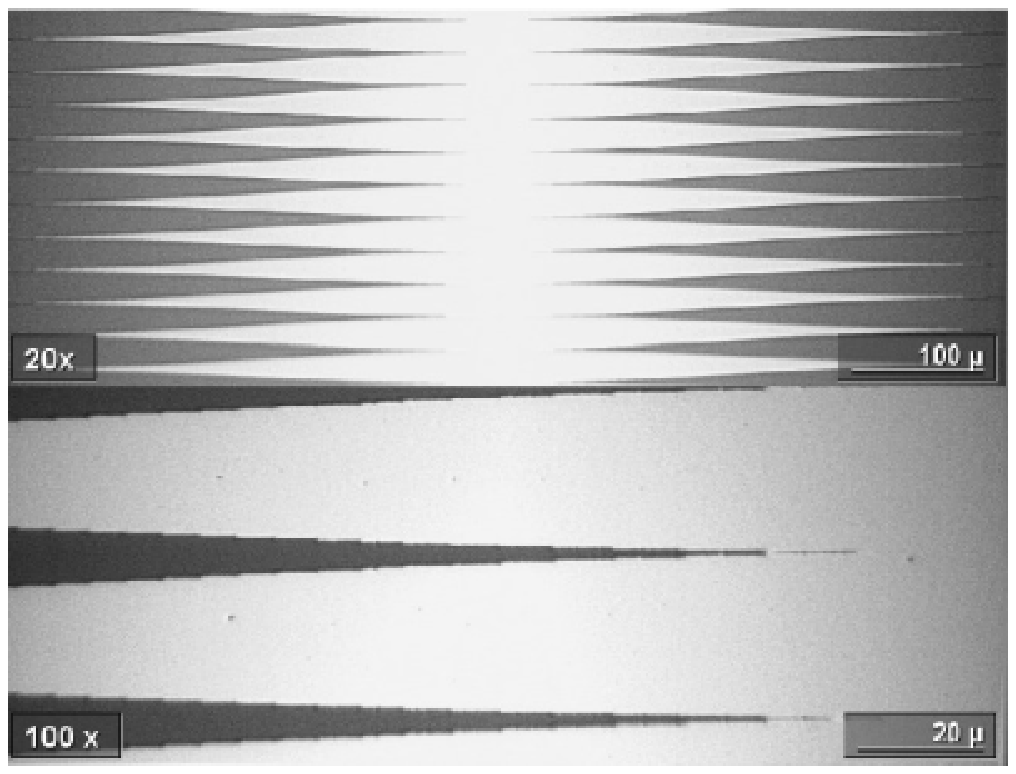}
\caption{\label{fig:mask} Optical microscope images of the final mask design. }
\end{figure}

The Leica EBPG5-HR  
EBL tool available at Nanofab can produce resist 
features down to a minimum size of $\sim$20~nm with  
a precision of $\pm$5~nm for high resolution resists.  The features are 
placed to an accuracy of $<$ 35~nm over a 125~mm writing area.
The EBL resist was developed on a commercially supplied quartz substrate, covered
by a layer of chromium that served as the opaque parts of the mask. 
Figure \ref{fig:mask} shows the final mask under 20x and 100x optical magnification.

The resolution used for the EBL tool was .25~$\micron$.  
Errors in the mask shape produce a leakage of 
light with an intensity of .25~$h^2$ where $h$ is the size of the error,
measured in
diffraction widths \citep{kuchner02b}.  Based on the EBL tool resolution, our mask should be capable of producing manufacturing-error-limited contrasts of  1.5$\times10^{-6}$ without the
use of an apodized Lyot stop; with our choice of experimental setup and wavelength, achieving a deeper contrast of 10$^{-8}$ at the peak would require 20 nm precision.  

\section{Modeling the Performance of the Mask}
\label{s2a}
One would like to know in advance how the mask responds to different sources of
error.  It is also instructive to try and reproduce the actual performance of 
the fabricated mask in an attempt to understand the major sources of error
in the experimental setup.  However, modeling the resulting diffraction
pattern requires the use of numerical methods such as Fast Fourier Transforms
(FFTs) which are not adept at accurately handling
simultaneous high resolution in both the imaging and pupil planes\citep{brigham}.

\begin{figure}
\plotone{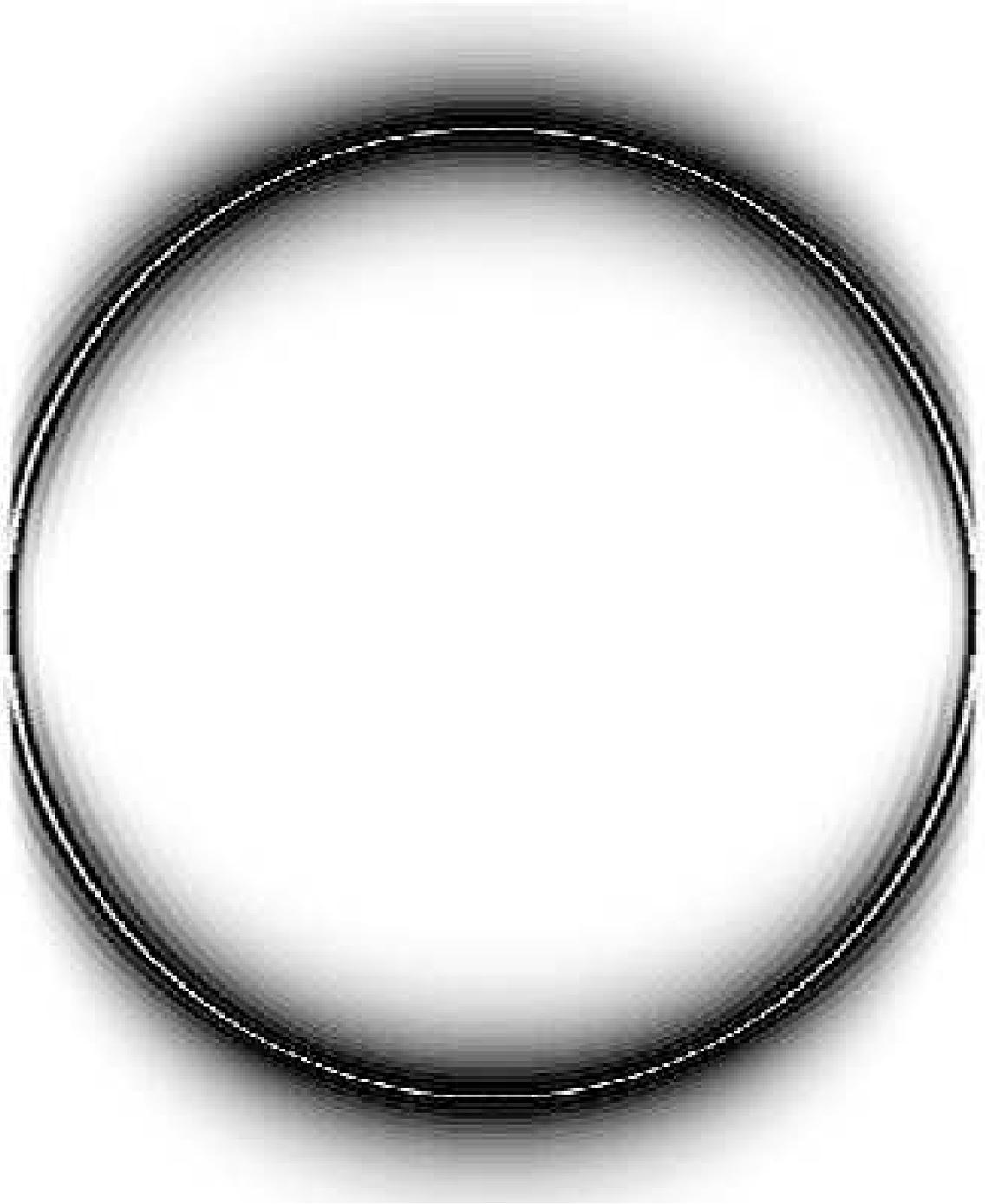}
\caption{\label{fig:modp} A model of the light distribution
in the pupil plane prior to a Lyot stop of a notch-filter mask design
with an $\epsilon$=.46, at the working wavelength of .633~\micron. }
\end{figure}

Modeling the resultant scalar diffraction pattern or point spread function (PSF) of 
an optical system can be summed up in the following combination of the wave
amplitude PSF of the original aperture $\hat{A}$, the mask function $\hat{M}$, and
the Fourier Transform of the Lyot Stop aperture function $\hat{L}$:
\begin{equation}
\hat{A^{\prime}}=\hat{A} \hat{M} \star \hat{L}
\end{equation}
where the star denotes convolution.  The numerical problem arises when
the arrays that represent $\hat{A}$ and $\hat{L}$ typically are Nyquist sampled,
corresponding to $\lambda/2D$, on order of the {\em maximum} scale of
the mask function.  To avoid this, we rely on the
fact that the Fourier Transform of $\hat{M}$ is semi-analytically described so 
that we can construct an accurate, coarsely sampled array that can then be 
used in the model.  The Fourier Transform of $\hat{M}, M(u,v_o)$ at a
particular value of $v_o$ is given by
\begin{equation}
M(u,v_o) = \int^{\frac{1}{2}}_{-\frac{1}{2}} \frac{\sin(\pi v_o \hat{M})}{\pi v_o} \exp(2\pi i u x) \mbox{d}x
\end{equation}  
In this way $M$ can be built up using a 1D FFT that is faster and more accurate
than its two dimensional analogue.  The final intensity at the image is 
given by $|\hat{A^{\prime}}\hat{A^{\prime \star}}|$.
Figure \ref{fig:modp} shows the light distribution before the Lyot Stop for the 
notch-filter mask we designed assuming no errors and a circular entrance pupil.
\section{Experiments and Results}
\label{s3}
The testbed used at Penn State was designed to test many different
coronagraph and shaped pupil designs.  We used a HeNe
laser as the light source and approximated a point source by placing a microscope objective lens in front of the
laser and placing the focused image on the
entrance aperture of a single mode fiber.  Light exiting
the fiber was collimated, followed by
a $\sim$3.16 mm entrance aperture.   An image was formed at the focal
plane image masks, which were mounted on a precision x-y-z stage.
The light was then re-collimated and passed through
a Lyot stop at an image of the entrance pupil.  Finally the light was
focused onto the CCD detector, where the final image was formed.  
The largest possible Lyot stop permitted by the linear mask design
we chose would have been shaped like the overlap region of two circles \citep{kuchner02}. Instead, for convenience we used
an iris as a Lyot stop.  The final diameter used for
the experiments was 1.8~mm.  This diameter is 75\% of what would be
expected theoretically.  We discuss the possible explanation for this in
Section \ref{s4}.

In order to measure a contrast ratio for any point in the PSF of the unblocked
point source or of the point source behind the mask, the counts in a particular
pixel must be normalized to the peak pixel counts of the unblocked point source
for a given exposure time.  Ideally one would measure the counts with the mask
out and in for the same exposure time.  The deepest
contrast one can achieve with this method is limited by the
nonlinearity of the CCD, in our case $\sim$20,000 ADU.  For a S/N $>$ 5 in a 
particular pixel,
 this level would correspond to a contrast of only $1.2\times10^{-3}$.  
To measure deeper contrasts one must take
longer exposures of the blocked point source and normalize the results
to an extrapolated count rate for the unblocked source.

To obtain reliable estimates of the count rates and fluxes of the different
configurations we observed the PSF over a range of linearly increasing exposure
times, taking care to avoid saturating the image.
We measured the peak pixel and total flux in each exposure
using the IRAF task IMEXAM.
We averaged the results and fit a linear model to them
using a least squares fitting routine in IDL
called LINFIT.  By extrapolating to a specific exposure time a normalization
for observations with the notch-filter mask could be obtained.
   
We repeated this procedure for the point
source through the quartz substrate of the notch-filter mask, with no substrate present, and with the Lyot stop completely open in order to determine the transmission of the
substrates and the throughput of the Lyot stop.
  Our reported results for the notch-filter mask are scaled to the
peak pixel count rate of the unblocked point source through the substrate.  We used the other measurements to gauge the throughput of the quartz substrate
 and the Lyot stop.

\begin{figure}
\plottwo{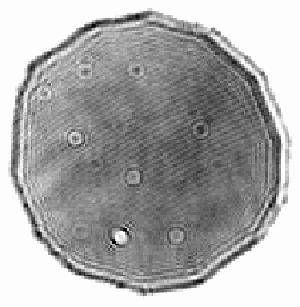}{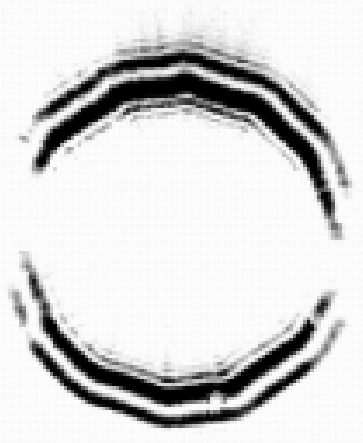}
\caption{\label{fig:pup} A comparison between pupil images
of the testbed with no mask (left) and with
a mask centered (right).}
\end{figure}

In the setup we also took observations of the pupil image both with and without the mask,
to gauge how well qualitatively the mask worked, compared to what is predicted by scalar
diffraction theory.  Figure \ref{fig:pup} shows that with the mask in place, the pupil
qualitatively resembles what is predicted by our model in Figure \ref{fig:modp}.  Note that both pupils in Figure \ref{fig:pup}
are at the same scale, and the bulk of the light falls outside the original pupil.

\begin{figure}
\epsscale{.5}
\plotone{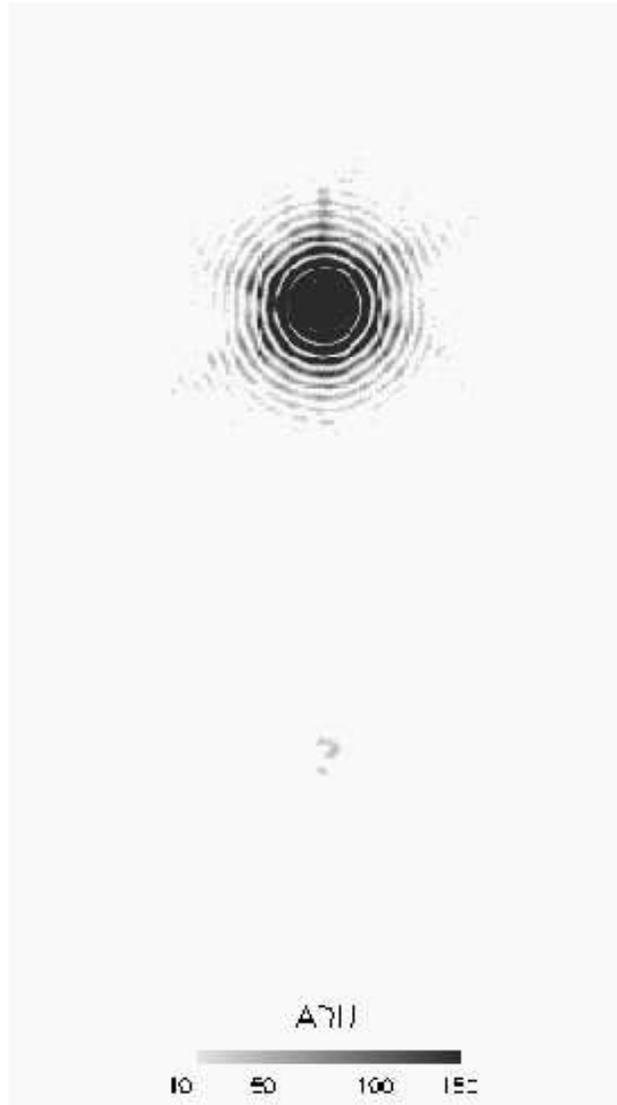}
\epsscale{1}
\caption{\label{fig:im} (top) 10 second image of the laser point source
 without the mask in place. (bottom) Image
with same intensity stretch with the mask in place.  The peak pixel value right
 image is 2.8$\times 10^{5}$ ADU.}
\end{figure}

Once the peak pixel count rates and fluxes were measured, deep observations of the
notch-filter mask were taken.  
Figure \ref{fig:im} shows two images set to the same contrast level and normalized
to the same exposure time.  The top image is one of the point source
without the notch-filter mask in place, and the bottom image is an 
exposure with the mask centered.  

The bar at the bottom
shows the number of counts on the detector associated with each level of the greyscale.  As can easily
be seen, the diffraction pattern of the light source is diminished greatly.

Figure \ref{fig:comp} demonstrates a more quantitative comparison with the mask
 present and absent and the undersized Lyot Stop in place for both 
configurations.  The 
figure shows the images taken during our experiments
 azimuthally averaged over all angles except for 20$^\circ$ on either side of 
the notch-filter's mask axis to avoid the region completely blocked by
the mask.  The curves are normalized to the peak pixel 
count rate of the unblocked image.  We have converted the spatial scale
in pixels to units of $\lambda/D$ by multiplying by a factor of 
$l_{pix}/(\lambda f_\#)$, where $l_{pix}$ is the width of a pixel
in \micron.  By doing this one can scale our results for 
existing or future telescopes.  It is important to note that this 
scale reflects
the degradation of spatial resolution due to a Lyot Stop that
is undersized.  The diffraction pattern is 
clearly suppressed by at least 2 orders of magnitude within
10$\lambda/D$, with a contrast of 9.5$\times10^{-6}$ reached at 3$\lambda/D$.  
In the course of our experiments we found that the mask was not completely 
opaque, allowing a small fraction of light transmission.  We measured 
the magnitude of this mask transmisivity (MT) as well with
the Lyot Stop in place and show it in 
Figure \ref{fig:comp} for comparison.  It appears that most of
the residual light corresponds to the wings of this transmission.

Table \ref{tab1} allows us to compare the relative throughput of the 
notch-filter mask to a setup without a Lyot stop.
We define throughput as the ratio of flux for a certain design to the flux of
the system with a completely open Lyot stop and no mask in place.  
We also measured the point source through the  
quartz substrate of the mask.  As can be seen, the throughput of the  notch-filter mask+Lyot stop combination is $\sim$27\%.

\section{Discussion}
\label{s4}
Our experiments did not attain the mask performance levels
expected from scalar diffraction theory.  In this section we will quantify the effect of some errors that degraded the contrast, and speculate on the
potential uses of this mask for space-based planet searches.

MT, the
finite size of the point source, mask alignment errors, and mask fabrication
errors all combine to explain the degraded performance of the notch 
filter mask.  These effects can be estimated and collected
into an error budget to guide further testing of the mask and drive
improvements in our setup.
 
The MT for dark parts of the mask should be less than
the contrast requirements. Degradation from light transmission
 can be estimated by
assuming a $\lambda/D$ by $\lambda/D$ hole in the mask with fractional 
transmission
$f$.  The central intensity of the leakage would be $\sim .25f$ as found in \citet{kuchner02}.  

The transmission
 flux measured in Table \ref{tab1} is $3\times10^{-3}$ times the unblocked
point source, 
giving a transmission peak intensity of 7.5 $\times10^{-4}$.  This is larger
than what is observed at the center, but one can estimate what would be
expected further away--the PSF is $\sim$10$^{-2}$ the peak at the first
Airy ring, which for the transmission gives an intensity of $\sim10^{-5}$, which is more consistent with what is seen further from the center.  The mask may not be uniformly transmissive and slightly thicker toward the center, which
could account for the suppression of the peak core. 
The opaque parts of the mask are covered by a 105 nm thick layer of chromium;
if this layer is doubled or tripled it will push the MT to $\sim10^{-8}$ 

If the error in alignment with respect to the mask is larger than the physical size of the
point source, then the leakage is $\sim (\Delta\theta/\theta_{1/2})^{4}$ where $\Delta\theta$
is the error in alignment and $\theta_{1/2}$ is the half power position
 of the mask \citep{kuchner02}.  The size of our single mode fiber core, 5 $\micron$,
ensures that the leakage due to it is $\ll$ the leakage due to misalignment of the
mask.
We have measured the half power of the mask to be $\sim$8 pixels or 192 $\micron$ in the
focal plane.  Our precision stage had an estimated accuracy of $\sim$16 $\micron$, based
on half the value of the smallest movement possible in the focal plane.  The
leakage would be $\sim4.8\times10^{-5}$ that of the unblocked point source.  

The surface roughness of our lenses will dictate the levels of scattered light we 
should observe and allow us to estimate the contribution of scattered light 
to the degradation in contrast.  We measured the surface roughness
of one of our lenses with a profilometer at Nanofab and obtained an estimate of the
RMS roughness (See \citet{elson1}).  Scattered 
light levels are $\propto$ $\delta_{rms}^2$,
 assuming a collection of plane gratings that diffract (scatter) light
into angles of interest.   
This formalism is for an opaque surface that reflects light.  However,
the results for a series of uncorrelated surfaces (i.e. an optical setup of many
lenses) give similar results provided that the roughness is separated on scales
$>$ $2\lambda$ \citep{elson2}. We find that 
the RMS roughness of the lens is .4 nm, which can be compared to the measured roughness of HST,
 $\sim$5.5 nm.  Therefore we estimate that the scattered light levels should
be $\sim(\delta/\delta_{HST})^2$ less than that of HST, corresponding to a contrast
level of $3\times10^{-7} (x/14.5)^{-2.19}$, where $x$ is in multiples of 
$\lambda/D$
\citep{bb1,malbet95}.  This corresponds to a scattered light level of $\sim$9$\times 10^{-6}$
at 3 $\lambda/D$ and $1\times 10^{-6}$ at $8\lambda/D$.
  More accurate measurements and
analysis are needed to better quantify the limitations of scattered light in the 
lab, as the above comparison is not necessarily accurate with such small scattering
angles \citep{bb1}.

\begin{figure}
\plotone{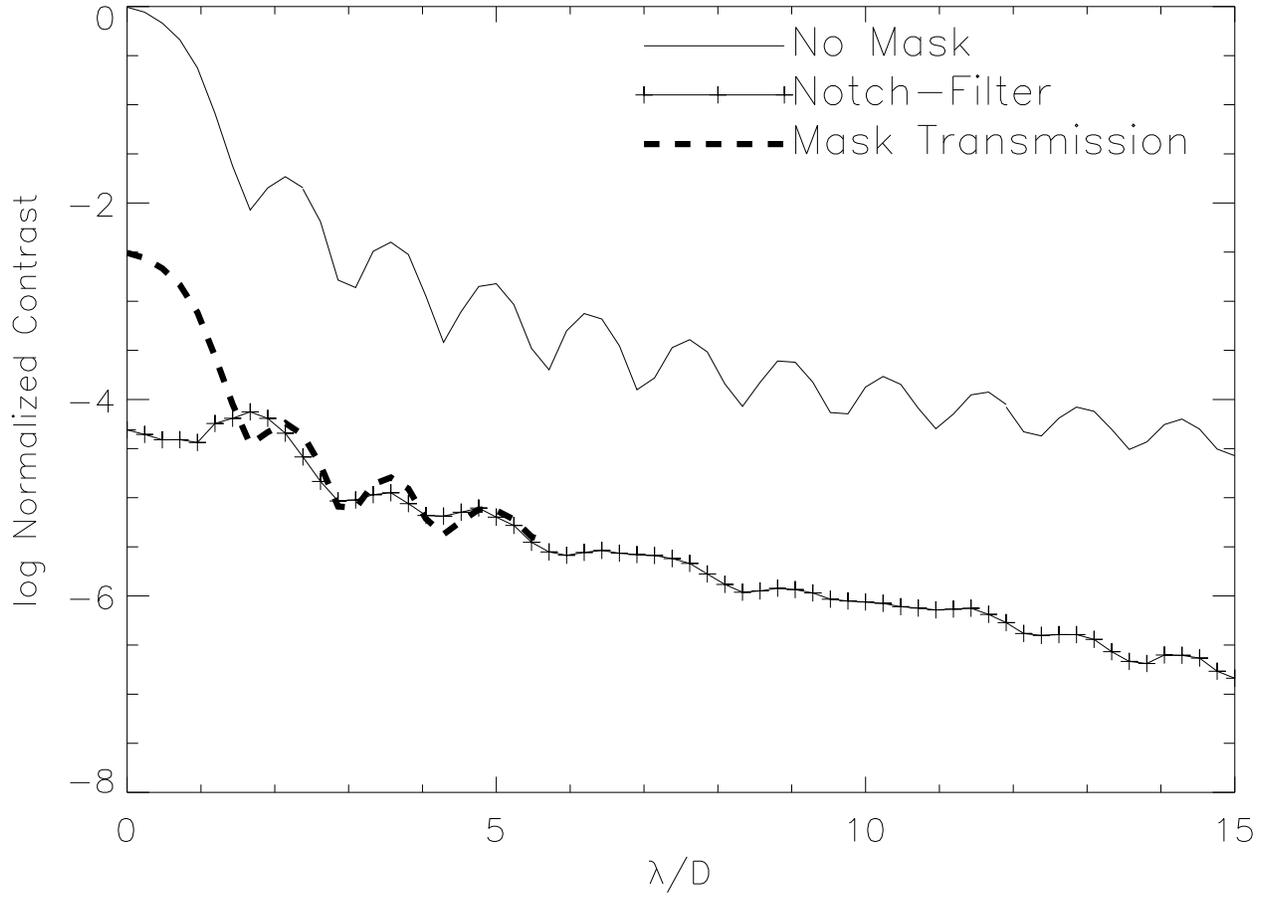}
\caption{\label{fig:comp} Comparison between the unblocked point source, the
notch-filter mask, and mask transmissivity.
  Each image was azimuthally averaged to within 20$^{\circ}$ of the notch 
filter mask axis. 
}
\end{figure}

MT and scattered light dominate the source of errors at $\sim$3$\lambda/D$, 
which is consistent with what is seen in Figure \ref{fig:comp}.  The resulting PSF
 with the notch-filter mask resembles the MT PSF, close to the PSF core, 
where the residual Airy Pattern of the MT dominates.  Further from the core
the Airy pattern is less distinct, most likely due to speckles from light 
scattered from the microroughness of our lenses.

A Lyot stop of diameter $\sim$2.4 mm should have sufficed for
a contrast of 10$^{-6}$.  Experimentally we found
that an undersized Lyot stop with 75\% the diameter of the theoretical
design appeared more efficacious.  This was based on an initial belief that the
degradation was caused by excess scattered light or slight misalignments
of the Lyot stop and the optical beam.  In those cases, undersizing the 
Lyot Stop would compensate for low levels of leakage.  However, since
it is apparent that the main cause of the degradation in contrast
is due to the MT, undersizing the stop simply reduces throughput. 

The design ``as-is'' already could have significant science benefits
in space.  Observations at the scattered
light limit of HST coupled with PSF subtraction
(shown to give an improvement of
contrast of around a factor of 50-100) could yield contrast levels of 
$\sim10^{-7}$ \citep{schneider,grady}.  For HST, the Lyot stop
would need to be designed such that the central obscuration and 
support pads would be adequately blocked at a cost in throughput.  The 
Lyot stop would be the overlap of three HST pupils, just as in the 
ideal case.  
  If we assume that
 with sufficient integration time we can reliably detect planets at this
 contrast level we can speculate how useful HST would be for
a planet search.
An instrument on HST optimized for coronagraphy 
could become a test bed for future
TPF coronagraph technology.    This setup would allow a limited extrasolar
planet direct imaging survey around nearby stars and white dwarfs.  As an 
example we consider our reported contrast at 3$\lambda/D$ in the J~band
on HST with PSF subtraction.
Given the best results one could expect $\Delta$J=17.5 and observe
 1~Gyr old 3~M$_{Jup}$ planets 10~AU from their host stars out to 30~pc and a 
10-100~Myr old 2 $M_{Jup}$ at 6.3 AU around
$\beta$ Pictoris \citep{burrows}.

\acknowledgements
This work is supported by the National Science
Foundation with grant AST-0138235, NASA grants NAG5-12115 and
NAG5-11427 and Ball Aerospace.  J.D. acknowledges funding by a NASA GSRP fellowship 
under grant NGT5-119. M.J.K. acknowledges the support of the Hubble Fellowship Program of the Space Telescope Science Institute.   Thanks also to Curtis DeWitt for his invaluable help in the 
lab, Deqing Ren for help with the experimental setup, and Dan McDavitt and Shane 
Miller for obtaining surface roughness scans.

\bibliography{apjbib}
\bibliographystyle{apj}

\clearpage

\clearpage

\begin{deluxetable}{ccc}
\tablecolumns{3}
\tablewidth{0pc}
\tablecaption{\label{tab1}Table of the flux and peak pixel count rate.  In each case
the flux is without a blocking filter present.}
\tablehead{
\colhead{Configuration} & \colhead{Flux} & \colhead{
 Peak Pixel Count Rate } \\
 & \colhead{(ADU s$^{-1}$ cm$^{-2}$)} & \colhead{(ADU s$^{-1})$} 
}
\startdata
Notch-filter glass & 7$\times 10^{8}$ & 2.2$\times10^{6}$ \\
No Mask & 8.0$\times10^{8}$ & 2.4$\times10^{6}$ \\
No Mask, No stop & 2.6$\times 10^{9}$ &  1.9$\times10^7$\\
Mask Transmissivity & 2.3$\times 10^{6}$ & 7.2$\times 10^{3}$ \\
\enddata
\end{deluxetable}

\clearpage

\end{document}